\documentclass[preprint,showpacs,preprintnumbers,amsmath,amssymb]{revtex4}

\usepackage{graphicx,subfigure,textcomp}
\usepackage{dcolumn}
\usepackage{bm}
\usepackage{mathrsfs,slashed}
\usepackage{natbib}

\begin{document}


\title{Finite volume corrections to the electromagnetic current of the nucleon}


\author{Ludwig Greil}
\affiliation{Institut f\"ur Theoretische Physik, Universit\"at Regensburg, D-93040 Regensburg, Germany}
\author{Thomas R. Hemmert}
\affiliation{Institut f\"ur Theoretische Physik, Universit\"at Regensburg, D-93040 Regensburg, Germany}
\author{Andreas Sch{\"a}fer}
\affiliation{Institut f\"ur Theoretische Physik, Universit\"at Regensburg, D-93040 Regensburg, Germany}
\date{\today}

\begin{abstract}
We compute corrections to both the isovector anomalous magnetic moment and the isovector electromagnetic current of the nucleon to $O(p^3)$ in the framework of covariant two-flavor Baryon Chiral Perturbation Theory. We then apply these corrections to lattice data for the anomalous magnetic moment from the LHPC, RBC \& UKQCD and QCDSF collaborations.
\end{abstract}

\maketitle

\section{\label{sec:Intro}Introduction}
One of the major challenges in nuclear physics today is to understand the structure of the nucleon arising from QCD dynamics. Major progress is made by the continuous improvement of lattice QCD, see e.g.~\cite{Hagler:2009ni}. However, in some cases results obtained from lattice simulations do not seem to extrapolate naturally to the experimentally known results. A prominent example are the lattice data for the electromagnetic form factors of the nucleon, a quantity for which high precision experimental data is available~\cite{Yamazaki:2009zq,Syritsyn:2009mx,Bratt:2010jn,Collins:2011mk}. It is rather plausible that this mismatch is due to the strong dependence on pion mass but quantitative numerical proof is still missing. \\The computation of observables on the lattice in general suffers from a number of systematic uncertainties: Both lattice spacing and lattice volume are finite, and simultaneously most simulations use quark masses that are much larger than the physical ones. Thus the quality of lattice results depends on the control over a threefold extrapolation: the continuum extrapolation ($a\rightarrow 0$), the extrapolation to the thermodynamic limit ($V\rightarrow 0$) and the chiral extrapolation ($m_q\rightarrow m_q^{\text{phys}}$).\\Therefore it is important to study all observables as functions of the quark masses or equivalently, as functions of the squared pseudoscalar mass. It is also desirable to be able to make a prediction of how large finite volume effects are for a certain observable. A suitable framework that allows one to predict pseudoscalar mass dependence as well as volume dependence for baryonic observables is called covariant Baryon Chiral Perturbation Theory (BChPT). It has been shown that BChPT provides good chiral extrapolations~\cite{Dorati:2007bk,Procura:2006gq} while controlling finite volume corrections~\cite{Gasser:1987zq,Hasenfratz1990241,AliKhan:2003cu,Procura:2006bj}.\\ In this paper we investigate the size of finite volume effects for nucleon form factor calculations. We work out the finite volume corrections to the anomalous magnetic moments of the nucleon to $O(p^3)$ in two-flavor BChPT. We then confront these corrections with lattice data. Using techniques presented in \cite{Sachrajda:2004mi} we also calculate the finite volume corrections to the electromagnetic current $\mathcal{J}^{\mu}$.
\section{\label{sec:IVAMM}Finite Volume Corrections to the isoscalar and isovector anomalous magnetic moment of the nucleon}
\begin{figure}[t]
\centering
\subfigure[][]{\includegraphics[]{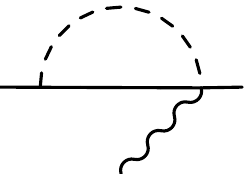}}\quad\subfigure[][]{\includegraphics[]{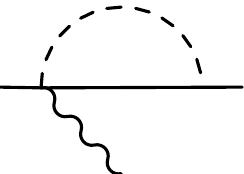}}\quad\subfigure[][]{\includegraphics[]{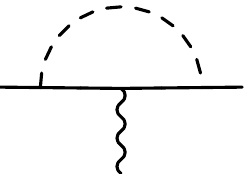}}\\
\subfigure[][]{\includegraphics[]{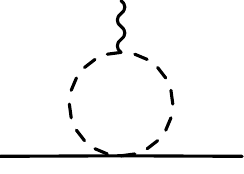}}\quad\subfigure[][]{\includegraphics[]{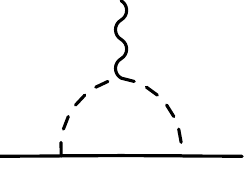}}\quad\subfigure[][]{\includegraphics[]{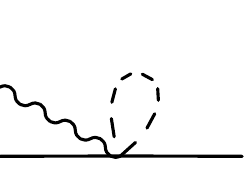}}
\caption{Feynman diagrams contributing to the anomalous magnetic moments $\kappa_v$ and $\kappa_s$ to $O(p^3)$ in BChPT. The solid line represents a nucleon, the dashed line denotes a pion and the wiggly line denotes a photon. All pion-nucleon and pion-nucleon-photon vertices stem from the leading order Lagrangian, where the pion-photon vertex is contained in $\mathscr{L}^{(2)}_{\pi\pi}$.}
\label{fig:diagrams}
\end{figure}

\begin{figure}
\includegraphics[width=0.6\textwidth]{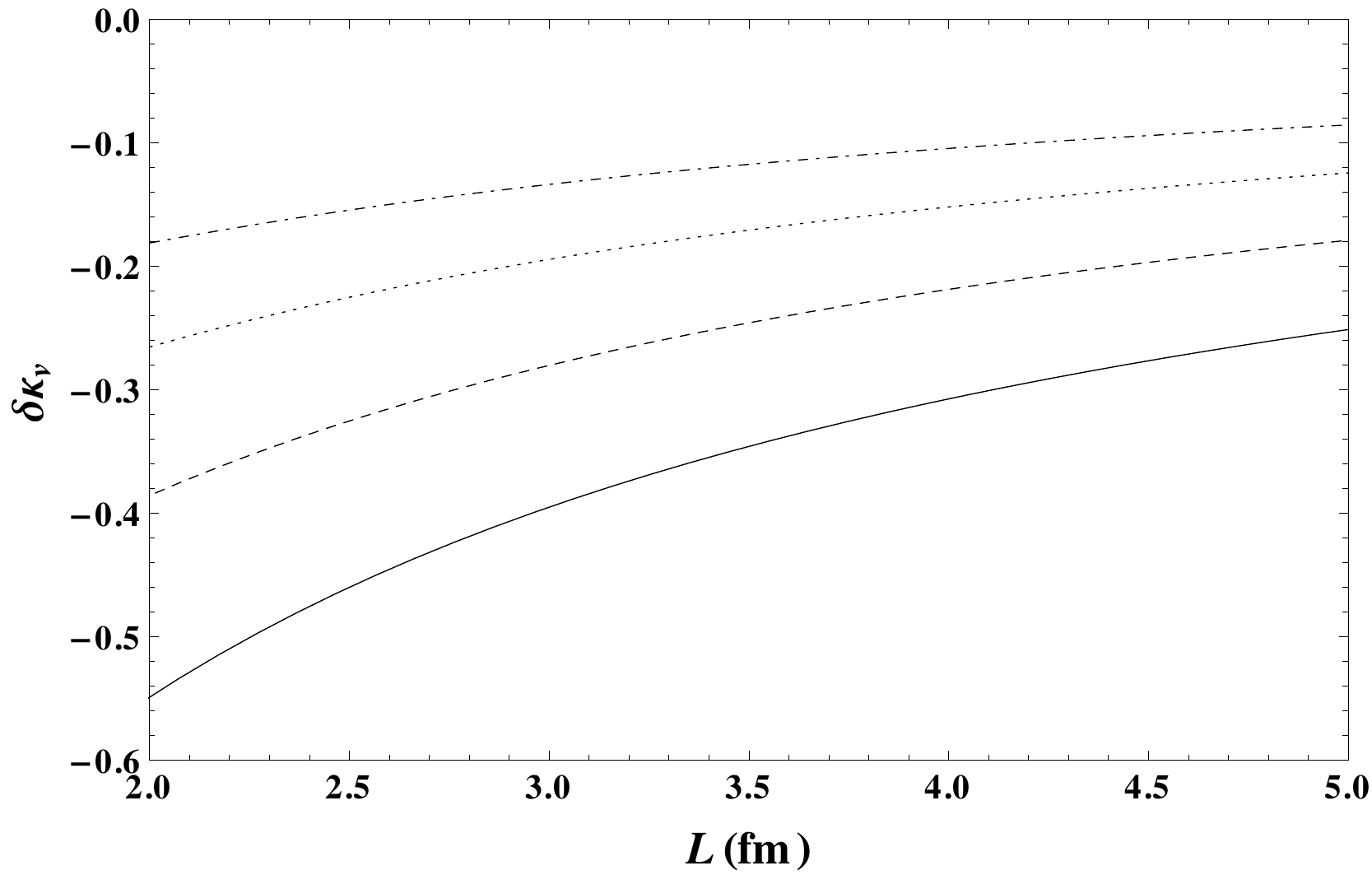}
\caption{The finite volume correction $\delta\kappa_v$ for different values of $m_{\pi}L$: $2$ (solid), $2.5$ (dashed), $3$ (dotted) and $3.5$ (dashdotted). We have the physical values for $f_{\pi}$, $g_A$ and $M_0\equiv M_N$, which is valid to the order we are working.}
\label{fig:constmpL}
\end{figure}

We employ $SU(2)_f$ BChPT as introduced in Ref.~\cite{Gasser:1987rb,Becher:1999he}. In this effective field theory, the elementary degrees of freedom are the pseudo Goldstone bosons (the pion fields) and the nucleon fields. In particular, we make use of the so-called modified infrared regularization scheme ($\overline{\text{IR}}$) as described in~\cite{Gail:2007,Dorati:2007bk}. This regularization scheme is a modification of the infrared regularization proposed in Ref.~\cite{Becher:1999he}.\\
To calculate the anomalous magnetic moment of the nucleon to $O(p^3)$, the effective Lagrangian
\begin{align}
\mathscr{L}_{\text{eff}}=\mathscr{L}_{\pi\pi}^{(2)}+\mathscr{L}_{N\pi}^{(1)}+\mathscr{L}_{N\pi}^{(2)}+\mathscr{L}_{N\pi}^{(3)},
\end{align}
is needed. The contributions $\mathscr{L}_{\pi\pi}^{(2)}$, $\mathscr{L}_{N\pi}^{(1-3)}$ have been taken from Refs.~\cite{Bernard:1995dp,Fettes:2000gb}. The Dirac and Pauli form factors of the nucleon $F_1(Q^2)$ and $F_2(Q^2)$ are defined as follows:
\begin{align}
\langle N(p^{\prime},s^{\prime})|\mathcal{J}^{\mu}|N(p,s)\rangle=\bar{u}(p^{\prime},s^{\prime})\left[\gamma^{\mu}F_1(Q^2)+i\sigma^{\mu\nu}\frac{q_{\nu}}{2M_N}F_2(Q^2)\right]u(p,s)\times\eta^{\dagger}\openone\eta,
\end{align}
where $\mathcal{J}^{\mu}=\frac{2}{3}\bar u\gamma^{\mu}u-\frac{1}{3}\bar d\gamma^{\mu}d$ denotes the electromagnetic current, $p$, $p^{\prime}$ are the initial and final nucleon momenta, $s$, $s^{\prime}$ are the spin vectors and the momentum transfer $q$ is defined as $q=p^{\prime}-p$. Furthermore, $Q^2=-q^2$ denotes the virtuality of the photon and $M_N$ is the nucleon mass ($M_N=939\,\text{MeV}$). Here, $\eta$ is a two-component vector describing the isospin content, i.e. $\eta=(1,0)^T$ represents a proton and $\eta=(0,1)^T$ represents a neutron. We have $F_1(0)=1$ for the proton because $\mathcal{J}$ is a conserved quantity and $F_2(0)$ measures the anomalous magnetic moment in nuclear magnetons. The nucleon form factors are related to the isoscalar and isovector form factors via
\begin{align}
F_{1,2}^{(s)}(Q^2)=F^{(p)}_{1,2}(Q^2)+F^{(n)}_{1,2}(Q^2),\qquad F_{1,2}^{(v)}(Q^2)=F^{(p)}_{1,2}(Q^2)-F^{(n)}_{1,2}(Q^2).
\end{align}
Thus, we find the following relations for the nucleon matrix elements of the electromagnetic current:
\begin{align}
\langle N(p^{\prime},s^{\prime})|\mathcal{J}^{\mu}_{(s)}|N(p,s)\rangle&=\bar{u}(p^{\prime},s^{\prime})\left[\gamma^{\mu}F_1^{(s)}(Q^2)+i\sigma^{\mu\nu}\frac{q_{\nu}}{2M_N}F_2^{(s)}(Q^2)\right]u(p,s)\times\eta^{\dagger}\frac{1}{2}\eta,\\
\langle N(p^{\prime},s^{\prime})|\mathcal{J}^{\mu}_{(v)}|N(p,s)\rangle&=\bar{u}(p^{\prime},s^{\prime})\left[\gamma^{\mu}F_1^{(v)}(Q^2)+i\sigma^{\mu\nu}\frac{q_{\nu}}{2M_N}F_2^{(v)}(Q^2)\right]u(p,s)\times\eta^{\dagger}\frac{\tau^3}{2}\eta.
\end{align}
Here, $\mathcal{J}_{(s)}^{\mu}$ and $\mathcal{J}_{(v)}^{\mu}$ denote the isoscalar and isovector currents. The isovector and isoscalar form factors are linear combinations of the quark form factors:
\begin{align}
F_{1,2}^{(v)}(Q^2)&=F_{1,2}^u(Q^2)-F_{1,2}^d(Q^2)\equiv F_{1,2}^{u-d}(Q^2),\\
F_{1,2}^{(s)}(Q^2)&=F_{1,2}^u(Q^2)+F_{1,2}^d(Q^2)\equiv F_{1,2}^{u+d}(Q^2).
\end{align}
Furthermore, we define $\kappa_v\equiv F_2^{(v)}(0)=3.71$ at physical $m_{\pi}$. For the isovector form factor, the disconnected diagrams, which are usually omitted in lattice simulations, cancel in the isospin limit, while for the isoscalar form factor, omitting them adds another source of uncertainty.\\
In this work, we calculate the finite volume corrections in two different ways: in this section, we assume that Lorentz invariance is still intact and we directly calculate the corrections for the anomalous magnetic moments. In sect. \ref{sec:FVCEMC} we assume Lorentz invariance is broken and thus, we have to calculate the corrections to the matrix elements of the electromagnetic current because these matrix elements can no longer be decomposed into two form factors.\\
The leading order corrections to the anomalous magnetic moments come from the Feynman diagrams shown in Fig.~\ref{fig:diagrams} and the finite volume corrections are calculated from the same Feynman diagrams. We do not impose any restrictions on the time direction, whereas for the spatial directions, we assume periodic boundary conditions.
If the spatial volume $L^3$ is not too small, the finite volume effects stem primarily from pions propagating around the spatial box. This leads to the so-called $p$-expansion, which is valid for small pion masses and large spatial volumes ($V=L^3$) such that $m_{\pi}L\gg 1$. To calculate the finite volume corrections from these Feynman diagrams, we make the following replacement when integrating over all possible loop momenta $k=(k_0,\mathbf{k})$:
\begin{align}
\int\frac{d^3\mathbf{k}}{(2\pi)^3}I(k_0,\mathbf{k})\rightarrow\frac{1}{L^3}\sum_{\mathbf{k}}I(k_0,\mathbf{k})-\int\frac{d^3\mathbf{k}}{(2\pi)^3}I(k_0,\mathbf{k}).
\end{align}
In the sum, $\mathbf{k}$ represents a discrete set of momenta that are allowed for a box with periodic boundary conditions and a spatial length $L$, i.e. $\mathbf{k}=2\pi(n_x,n_y,n_z)^T/L$, where $n_x$, $n_y$ and $n_z$ are integer numbers. Using Lorentz decomposition, all appearing tensor integrals can be reduced to scalar integrals (see Appendix~\ref{app:BI}). Using the definitions
\begin{align}
H_{10}(L)&=\int\frac{dk_0}{2\pi}\left(\frac{1}{L^3}\sum_{\mathbf{k}}-\int\frac{d^3\mathbf{k}}{(2\pi)^3}\right)\frac{1}{m^2-k^2-i\epsilon},\\
H_{01}(L)&=\int\frac{dk_0}{2\pi}\left(\frac{1}{L^3}\sum_{\mathbf{k}}-\int\frac{d^3\mathbf{k}}{(2\pi)^3}\right)\frac{1}{M^2-k^2-i\epsilon},\\
H_{11}^{(0)}(L)&=\int\frac{dk_0}{2\pi}\left(\frac{1}{L^3}\sum_{\mathbf{k}}-\int\frac{d^3\mathbf{k}}{(2\pi)^3}\right)\frac{1}{\left[m^2-k^2-i\epsilon\right]\left[M^2-(p-k)^2-i\epsilon\right]},
\end{align}
where $p$ represents the nucleon four momentum, the $O(p^3)$ result for the finite volume corrections can be written as follows:
\begin{align}
\begin{split}
\delta\kappa_s^{(3)}(m_{\pi},L)&=\kappa^{(3)}_s(L)-\kappa^{(3)}_s(\infty)\\&=-\frac{2M^2g_A^2}{f_{\pi}^2}\left\{\left(\frac{\partial}{\partial M^2}+\frac{m^2}{M^2}\frac{\partial}{\partial M^2}-\frac{1}{M^2}\right)H_{01}(L)\right.\\
&\quad\left.+m^2\left(\frac{2}{M^2}-\frac{m^2}{M^2}\frac{\partial}{\partial M^2}+\frac{\partial}{\partial M^2}\right)H_{11}^{(0)}(L)+\frac{1}{M^2}H_{10}(L)\right\},
\end{split}\\
\begin{split}
\delta\kappa_v^{(3)}(m_{\pi},L)&=\kappa^{(3)}_v(L)-\kappa^{(3)}_v(\infty)\\
&=\frac{2g_A^2}{3f_{\pi}^2}\left\{\left(M^2\frac{\partial}{\partial M^2}+m^2\frac{\partial}{\partial M^2}-1\right)H_{01}(L)\right.\\
&\quad\left.+m^2\left(2-m^2\frac{\partial}{\partial M^2}+M^2\frac{\partial}{\partial M^2}\right)H_{11}^{(0)}(L)+H_{10}(L)\right\}\label{eq:dkvfi}\\
&\quad-\frac{8g_A^2}{3f_{\pi}^2}\left\{\left(1+m^2\frac{\partial}{\partial m^2}\right)H_{10}(L)-H_{01}(L)\right.\\
&\quad\left.+\left(2m^2+m^4\frac{\partial}{\partial m^2}-M^2-m^2M^2\frac{\partial}{\partial m^2}\right)H_{11}^{(0)}(L)\right\}.
\end{split}
\end{align}
Here, $g_A$ and $f_{\pi}$ should be taken in the chiral limit, but to the order these corrections have been calculated, using the physical values is accurate. Note that (other than $L$) no new parameters have been introduced. For the form of the finite volume functions see Appendix~\ref{app:FVF}.\\
At this point, we would like to comment on the use of the simplified fit ansatz $\delta\kappa_v(m_{\pi},L)=a\exp(-m_{\pi}L)$ when trying to determine the infinite volume limit, as has been done in \cite{Collins:2011mk}. This form is motivated by finite volume meson ChPT, i.e. terms of the form \eqref{eq:fvol1}, in the limit of large $m_{\pi}L$ and only taking into account the $n=1$ contribution. If correct, one would expect to find one single value for the correction along the lines of constant $m_{\pi}L$. For the full expression for $\delta\kappa_V$, one finds, however, that the corrections are not constant along the lines of constant $m_{\pi}L$, see fig. \ref{fig:constmpL}. Most likely this behaviour is caused by the appearance of the mass scale $M_0$ in BChPT. Thus, we would advise to always use these $\mathcal{O}(p^3)$ BChPT formulae instead.
\section{\label{sec:LQCD}Comparison with results from Lattice QCD}
\begin{figure}[h]
\centering
\subfigure[][]{\label{fig:fitressetI}\includegraphics[width=0.6\textwidth]{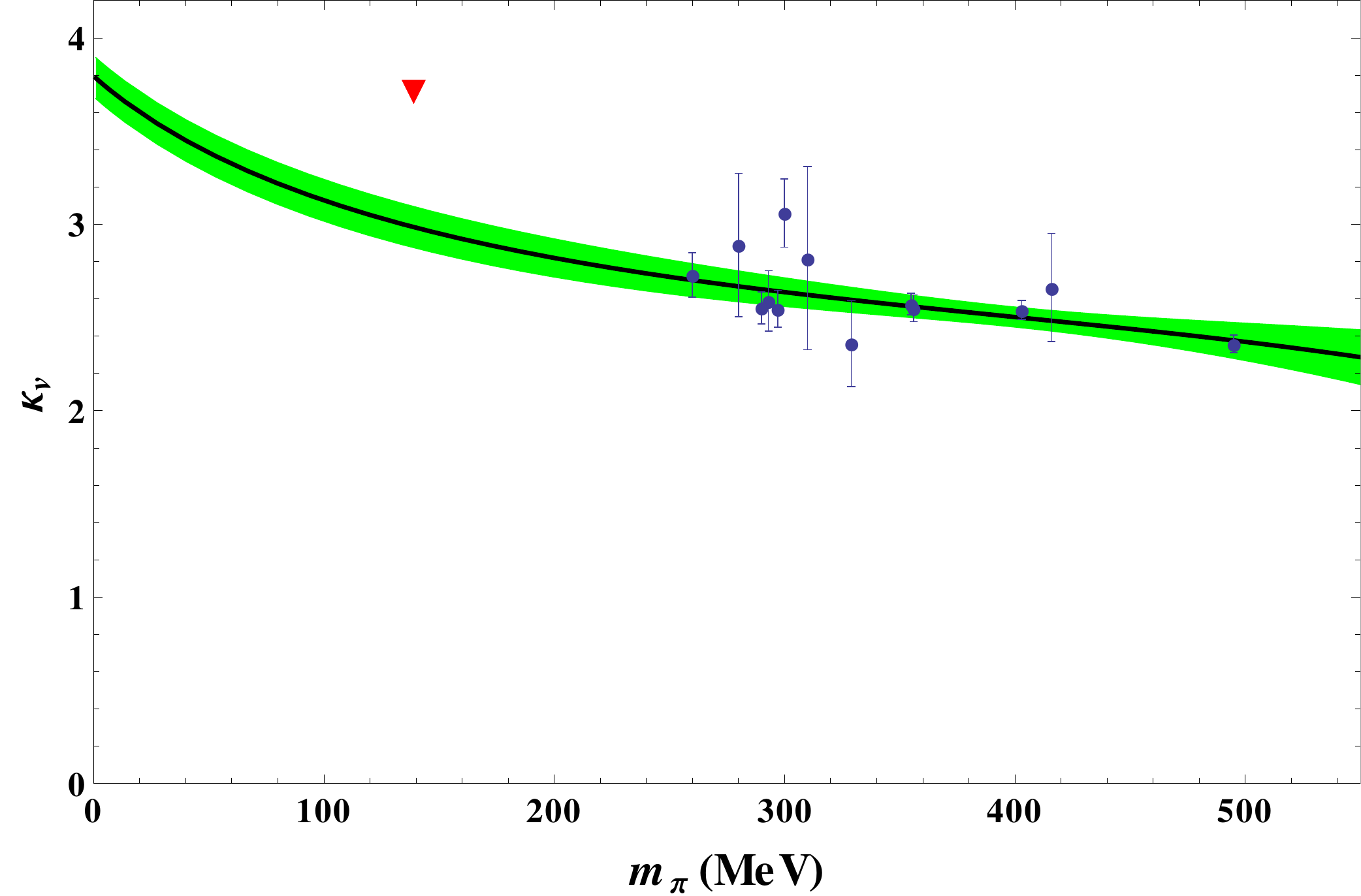}}\\
\subfigure[][]{\label{fig:fitressetII}\includegraphics[width=0.6\textwidth]{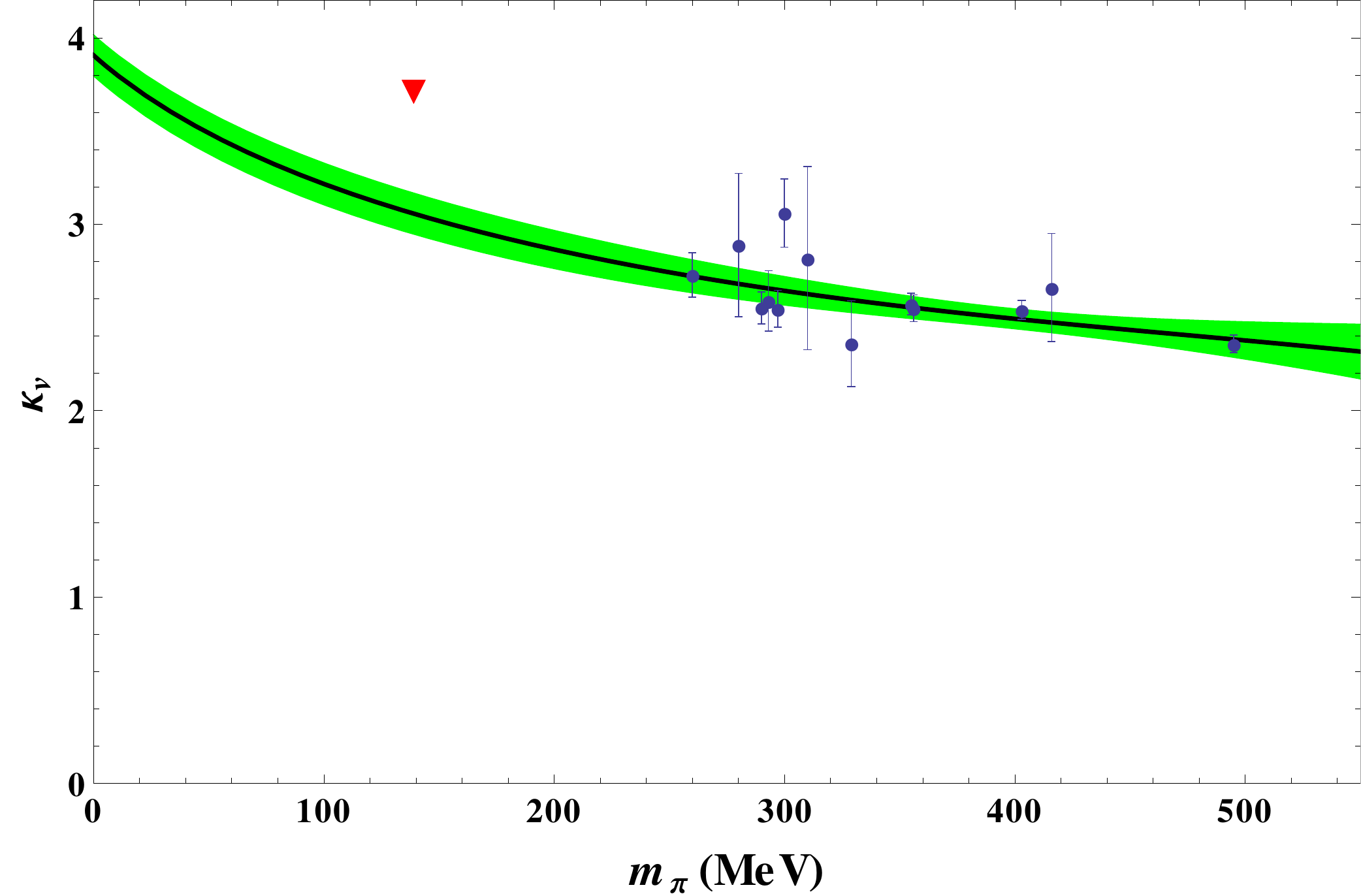}}
\caption{Fit results for two different sets of input parameters. The shaded area represents the $95\%$ confidence level band. Fig. \ref{fig:fitressetI} corresponds to the LECs labeled Set I in tab. \ref{tab:IPmass}, whereas for fig. \ref{fig:fitressetII}, Set II was used.}
\label{fig:fitres}
\end{figure}

Using the input parameters given in Table~\ref{tab:IPP} we can calculate the infinite volume quantities from the raw lattice data by first rewriting $\kappa_v$, given in lattice magnetons (short l.m.), in terms of nuclear magnetons (short n.m.) via the equation~\cite{Gockeler:2003jfa}
\begin{align}
\kappa_v^{\text{norm}}=\kappa_v^{\text{lattice}}\frac{M_N^{\text{phys}}}{M_N^{\text{lattice}}},
\end{align}
and then subtracting the finite volume corrections from this normalized quantity:
\begin{align}
\kappa_v^{\text{norm}}(m_{\pi},\infty)=\kappa_v^{\text{norm}}(m_{\pi},L)-\delta\kappa_v(m_{\pi},L).
\end{align}
In Table~\ref{tab:IPP} we present the input values for the numerical evaluation of the finite volume corrections. Since the corrections are proportional to a factor $(g_A/f_{\pi})^2$, choosing the physical values for $g_A$ and $f_{\pi}$ instead of the values in the chiral limit amounts to a mere multiplication with a common factor. Choosing the physical values for the pion decay constant and the axial coupling constant would result in multiplying the corrections with a factor of $0.99$. Thus, we use the physical values for both $g_A$ and $f_{\pi}$, because they are known more accurately and it does not change the corrected values $\kappa_v^{\text{corr}}$ significantly. From Table~\ref{tab:SUM} we can clearly conclude that the finite volume corrections are considerably large although $m_{\pi}L\approx3.8$ for the smallest pion masses. This set of finite volume corrected isovector anomalous magnetic moments can be used as input for the second fit.\\
The $O(p^4)$ BChPT formula for the anomalous magnetic moment is presented in Ref.~\cite{Kubis:2000zd,Gail:2007} and is of the following form:
\begin{align}
\begin{split}\label{eq:BChPTkv}
\kappa_v&=\frac{M_N}{M_0}\left[c_6-16M_0m_{\pi}^3e_{106}^r(\lambda)+\kappa_v^{(3)}+\kappa_v^{(4)}+O(p^5)\right],
\end{split}\\
\begin{split}
\kappa_v^{(3)}&=\frac{g_A^2m_{\pi}^2M_0}{8\pi^2f_{\pi}^2M^3}\left[(3m_{\pi}^2-7M^2)\log{\left(\frac{m_{\pi}}{M}\right)}-3M^2\right]\\
&\quad-\frac{g_A^2m_{\pi}M_0}{8\pi^2f_{\pi}^2M^3\sqrt{4M^2-m_{\pi}^2}}\left[3m_{\pi}^4-13m_{\pi}^2M^2+8M^4\right]\arccos{\left(\frac{m_{\pi}}{2M}\right)},
\end{split}\\
\begin{split}
\kappa_v^{(4)}&=-\frac{m_{\pi}^2}{32\pi^2f_{\pi}^2M_0^2}\left[4g_A^2(c_6+1)M_0^2-g_A^2(5c_5m_{\pi}^2+28M_0^2)\log{\left(\frac{m_{\pi}}{M_0}\right)}\right.\\
&\left.\quad+4M_0^2(2c_6g_A^2+7g_A^2+c_6-4c_4M_0)\log{\left(\frac{m_{\pi}}{\lambda}\right)}\right]\\
&\quad-\frac{g_A^2c_6m_{\pi}^3}{32\pi^2f_{\pi}^2M_0\sqrt{4M_0^2-m_{\pi}^2}}(5m_{\pi}^2-16M_0^2)\arccos{\left(\frac{m_{\pi}}{2M_0}\right)}.
\end{split}
\end{align}
Note that we distinguish between $M_0$, which is the nucleon mass in the chiral limit, and $M=M(m_{\pi})$, the running nucleon mass which appears in the $O(p^3)$ part of the whole $O(p^4)$ result. As input for the running nucleon mass we have chosen two different sets of LECs which are presented in Tab. \ref{tab:IPmass}. For the fitting procedures, we have only used $\kappa_v$ values for pion masses smaller than $500\,\text{MeV}$, which we consider to be a reasonable fitting window, considering that the upper bound estimates for the applicability of BChPT found in the literature range from pion masses of $350\,\text{MeV}$ to $600\,\text{MeV}$ \cite{Djukanovic:2006xc,Schindler:2007dr,Bernard:2006te}, depending on the observables considered. The results of these fits are presented in fig. \ref{fig:fitres}.
As one can see, the finite volume corrected lattice data does not naturally extrapolate to the known value at the physical point. This is most likely due to a lack of information in the region where one expects a lot of curvature (notice that in the interval $300\,\text{MeV}<m_{\pi}<500\,\text{MeV}$ all values are around $\kappa_v\approx2.4$ within errorbars).
\section{\label{sec:FVCEMC}Finite Volume Corrections to the electromagnetic current of the nucleon}
For the calculation of the finite volume corrections to the anomalous magnetic moments we have assumed Lorentz invariance to still be intact. But since we are doing all calculations in a box of finite spatial volume, Lorentz invariance is explicitly broken and thus, Lorentz decomposition of tensorial integrals is no longer possible, see Appendix~\ref{app:FVF}. A recalculation of the Feynman diagrams shown in Fig.~\ref{fig:diagrams} in a gauge where the polarization of the photon is perpendicular to its four momentum ($\varepsilon\cdot q=0$) shows that a decomposition of the matrix elements into only two different form factors is no longer possible, in accordance with what one would expect for broken Lorentz invariance.\\
Using Eq.~\eqref{eq:corrcurr}, one can calculate the corrections to the nucleon matrix element of the electromagnetic current for various momenta $p$ and $q=(q_0,\mathbf{q})$. All formulae calculated with BChPT are applicable for small $q^2$ (in Ref.~\cite{Bernard:1998gv} it was suggested that $|q^2|\leq0.1\,\text{GeV}^2$). We choose the lowest possible momentum transfers, i.e. $\mathbf{q}\in2\pi/L\{(1,0,0)^T,(0,1,0)^T,(0,0,1)^T\}$. Furthermore, we only consider spin projections along the $z$-axis. We then compare the matrix elements of the corrections to the infinite volume matrix elements using the $O(p^3)$ BChPT results calculated in~\cite{Kubis:2000zd,Gail:2007}. These take the form
\begin{align}\label{eq:ffv}
 F_1^{(v)}(q^2)=1+q^2\rho_1^{(v)}+O(q^4), \qquad F_2^{(v)}(q^2)=\kappa_v+q^2\rho_2^{(v)}+O(q^4),
\end{align}
where the $O(p^3)$ results for the slopes $\rho_1^{(v)}$ and $\rho_2^{(v)}$ are of the following form:
\begin{align}
\begin{split}
\rho_1^{(v)}&=B_{c1}(\lambda)-\frac{1}{96\pi^2f_{\pi}^2M_0^4}\biggl[7g_A^2M_0^4+2(5g_A^2+1)M_0^4\log\left(\frac{m_{\pi}}{\lambda}\right)+M_0^4\\
&\quad-15g_A^2m_{\pi}^2M_0^2+g_A^2m_{\pi}^2(15m_{\pi}^2-44M_0^2)\log\left(\frac{m_{\pi}}{M_0}\right)\biggr]\\
&\quad+\frac{g_A^2m_{\pi}}{96\pi^2f_{\pi}^2M_0^4\sqrt{4M_0^2-m_{\pi}^2}}\left[15m_{\pi}^4-74m_{\pi}^2M_0^2+70M_0^4\right]\arccos\left(\frac{m_{\pi}}{2M_0}\right),
\end{split}\\
\begin{split}
\rho_2^{(v)}&=B_{c2}(\lambda)+\frac{g_A^2}{96\pi^2f_{\pi}^2M_0^4(m_{\pi}^2-4M_0^2)}\biggl[-124M_0^6+105m_{\pi}^2M_0^4-18m_{\pi}^4M_0^2\\
&\quad+6(3m_{\pi}^6-22m_{\pi}^4M_0^2+44m_{\pi}^2M_0^4-16M_0^6)\log\left(\frac{m_{\pi}}{M_0}\right)\biggr]\\
&\quad+\frac{g_A^2}{48\pi^2f_{\pi}^2M_0^4m_{\pi}(4M_0^2-m_{\pi}^2)^{3/2}}\biggl[9m_{\pi}^8-84m_{\pi}^6M_0^2+246m_{\pi}^4M_0^4-216m_{\pi}^2M_0^6\\
&\quad+16M_0^8\biggr]\arccos\left(\frac{m_{\pi}}{2M_0}\right).
\end{split}
\end{align}
To arrive at $q^2\leq0.1\,\text{GeV}^2$ we choose the spatial length of the box to be $L=3.87\,\text{fm}$ which is significantly larger than the volumes available in current lattice simulations. For the numerical evaluation we truncate the infinite sum at $|n_i|=3$ and we have used the low energy constants tabulated in Tab.~\ref{tab:IPP}, which are taken from~\cite{Gail:2007}. We have chosen the Dirac representation and have normalized the Dirac spinors to $\bar{u}(p,s)u(p,s)=1$. For the numerical evaluation of the current matrix elements where we utilize the $O(p^3)$ BChPT results, Eqs. \eqref{eq:ffv} and \eqref{eq:corrcurr}, we have used the physical values for $g_A$, $f_{\pi}$ instead of the values in the chiral limit and have also set $M_0=M_N^{\text{phys}}$, which is accurate to the chiral order all these quantities have been calculated in. This then yields the results collected in tab.~\ref{tab:currres1}. They show clearly that to extract $F_1(q^2)$ and $F_2(q^2)$ one first has to correct the finite volume results. Again, one finds that the finite volume corrections to the matrix element of the electromagnetic current can be of $O(10\%)$, even for rather large box lengths $L$.
\section{\label{sec:CONC}Conclusion}
The results of our work can be summarized as follows:
\begin{enumerate}
\item{We have calculated the finite volume corrections to the anomalous magnetic moments $\kappa_{s,v}$ of the nucleon in the framework of $SU(2)_f$ covariant BChPT assuming Lorentz invariance is still intact.}
\item{We have found these corrections to be of $O(10\%)$ for pion masses smaller than $m_{\pi}<300\,\text{MeV}$ and box lengths of $L\approx2.5\,\text{fm}$.}
\item{The chiral extrapolation of the finite volume corrected lattice data for $\kappa_v$ still does not yield the experimentally known value. This is most likely due to the lack of information in the region $138\,\text{MeV}<m_{\pi}<250\,\text{MeV}$. Partly, these discrepancies might also originate from simulating two instead of three dynamical quark flavors.}
\item{When working in a finite volume, Lorentz invariance is broken. Therefore, we have redone all calculations without making use of Lorentz decomposition when dealing with tensorial loop integrals. This leads to an expression for the correction to the nucleon matrix elements of the electromagnetic current.}
\item{The uncorrected finite volume matrix elements can no longer be decomposed into the Pauli and Dirac form factors (see tab. \ref{tab:currres1}). These corrections are considerably large when choosing $m_{\pi}=m_{\pi}^{\text{phys}}$ and $|q^2|\approx0.1\,\text{GeV}^2$, i.e. $L=3.8\,\text{fm}$.}
\item{The problem that the lattice data does not naturally extrapolate to the experimentally known value for $\kappa_v$ can not be solved by taking finite volume corrections into account. However, having data for pion masses smaller than $250\,\text{MeV}$ and then applying these finite volume corrections should yield a reasonable result.}
\end{enumerate}
\acknowledgments{The authors would like to thank Andr\'{e} Sternbeck for providing the input parameters. We would also thank Peter C. Bruns for discussion. This work was supported by the Deutsche Forschungsgemeinschaft
SFB/Transregio 55.}
\newpage
\appendix
\section{\label{app:TAB}Tables}
\begin{table}[h]
\caption{Lattice data for both $M_N$ in MeV and $\kappa_v$ in terms of normalized physical magnetons from~\cite{Yamazaki:2009zq},~\cite{Syritsyn:2009mx} and~\cite{Bratt:2010jn}. Also shown are the $O(p^3)$ finite volume corrected values $\kappa_{v}^{\text{corr}}$. To calculate $\kappa_{v}^{\text{corr}}$, we have truncated the infinite sums appearing in \eqref{eq:dkvfi} at $n=20$.}
\begin{ruledtabular}
\begin{tabular}{c c c c c c}
Ref. & $L\,[\text{fm}]$ & $m_{\pi}\,[\text{MeV}]$ & $M_{N}\,[\text{MeV}]$ & $\kappa_v\,\left[\text{n.m.}\right]$ & $\kappa_{v}^{\text{corr}}\,\left[\text{n.m.}\right]$\\
\hline
\cite{Yamazaki:2009zq} & 2.7 & 329 & 1154(7) & 2.29(23) & 2.37(23)\\
                       & 2.7 & 416 & 1216(7) & 2.63(29) & 2.66(29)\\
\hline
\cite{Syritsyn:2009mx} & 2.688 & 297 & 1109(21) & 2.447(99) & 2.554(99)\\
                       & 2.688 & 335 & 1172(21) & 2.518(57) & 2.576(57)\\
                       & 2.688 & 403 & 1221(21) & 2.508(51) & 2.543(51)\\
\hline
\cite{Bratt:2010jn}    & 2.5 & 293 & 1107.1(111) & 2.456(162) & 2.602(162)\\
                       & 2.5 & 356 & 1154.8(80)  & 2.475(72)  & 2.554(72)\\
                       & 2.5 & 495 & 1288.4(80)  & 2.339(47)  & 2.359(47)\\
\hline
\cite{Collins:2011mk}  & 2.9 & 260 & & 2.608(115) & 2.728(115)\\
                       & 1.9 & 280 & & 2.505(311) & 2.889(311)\\
                       & 2.9 & 290 & & 2.466(80)  & 2.552(80)\\
                       & 2.3 & 300 & & 2.877(184) & 3.060(184)\\
                       & 1.7 & 310 & & 2.396(491) & 2.818(491)
\end{tabular}
\end{ruledtabular}
\label{tab:SUM}
\end{table}
\begin{table}[h]
\centering
\caption{Low energy constants used as input parameters for the fitting procedure taken at a renormalization scale $\lambda=1\,\text{GeV}$. The two sets parameters $c_2$, $c_3$, $c_4$ are taken from \cite{Becher:2001hv} while $M_0$, $c_1$ and $e_1^r$ were provided by \cite{Sternbeck2011}.}
\begin{ruledtabular}
\begin{tabular}{l c c c c c c}
      & $M_0\,[\text{GeV}]$ & $c_1\,[\text{GeV}^{-1}]$ & $c_2\,[\text{GeV}^{-1}]$ & $c_3\,[\text{GeV}^{-1}]$ & $c_4\,[\text{GeV}^{-1}]$ & $e_1^r(\lambda)\,[\text{GeV}^{-3}]$\\
\hline
Set I  & $0.893$ & $-0.757$ & $1.71$ & $-3.64$ & $2.13$ & $2.25$\\
Set II & $0.891$ & $-0.825$ & $2.66$ & $-4.50$ & $2.45$ & $2.22$\\
\end{tabular}
\end{ruledtabular}
\label{tab:IPmass}
\end{table}
\begin{table}[h]
\centering
\caption{Low energy constants used as input parameters for the numerical evaluations.}
\begin{ruledtabular}
\begin{tabular}{c c c c}
$g_A^0$ & $f_{\pi}\,[\text{MeV}]$ & $B_{c1}(M_0)\,[\text{GeV}^{-2}]$ & $B_{c2}(M_0)\,[\text{GeV}^{-2}]$\\
\hline
$1.2695$ & $92.4$ & $-1.3$ & $4.47$\\
\end{tabular}
\end{ruledtabular}
\label{tab:IPP}
\end{table}
\begin{table}
\centering
\caption{Low energy constants determined from the fit to the pion mass dependent anomalous magnetic moment using the $O(p^4)$ BChPT formula~\eqref{eq:BChPTkv}. All values have been obtained at $\lambda=1\,\text{GeV}$. The first row shows the values and the statistical errors obtained by the fit routine whereas the second row shows the values suggested to achieve a $95\%$ confidence level. }
\begin{ruledtabular}
\begin{tabular}{l c c c}
Input & $c_6$ & $e_{106}^r(\lambda)\,[\text{GeV}^{-3}]$ & $\chi^2/\text{d.o.f.}$\\
\hline
Set I & 3.61(5)  & 0.27(4) & 0.97\\
      & 3.61(11) & 0.27(8) & \\
\hline
Set II & 3.71(5)  & 0.22(4) & 1.03\\
      & 3.71(11) & 0.22(9)& \\
\end{tabular}
\end{ruledtabular}
\label{tab:lecresII}
\end{table}\newpage
\begin{table}
\centering
\caption{Numerical results for the current matrix element, the matrix element of the corrections and the matrix element of the electromagnetic current in finite volume. The isospin part of the matrix element has been omitted.}
\begin{ruledtabular}
\begin{tabular}{c c c c}
$\frac{L}{2\pi}\mathbf{q}$ &  $\langle N\left(p,\frac{1}{2}\right)|\mathcal{J}^{\mu}|N\left(p,\frac{1}{2}\right)\rangle$ & $\langle N\left(p,\frac{1}{2}\right)|\delta\mathcal{J}^{\mu}|N\left(p,\frac{1}{2}\right)\rangle$ & $\langle N\left(p,\frac{1}{2}\right)|\mathcal{J}^{\mu}(L)|N\left(p,\frac{1}{2}\right)\rangle$\\
\hline
$(1,0,0)^T$ & $\begin{pmatrix}0.765697\\0\\-0.604959i\\0\end{pmatrix}$ & $\begin{pmatrix}-0.0855116\\0.0380878\\0.0738641-0.00741085i\\0.0738641\end{pmatrix}$ & $\begin{pmatrix}0.680185\\0.0380878\\0.0738641-0.612369i\\0.0738641\end{pmatrix}$\\
\hline
$(0,1,0)^T$ & $\begin{pmatrix}0.765697\\0.604959i\\0\\0\end{pmatrix}$ & $\begin{pmatrix}-0.0855116\\0.0738641+0.00741085i\\0.0380878\\0.0738641\end{pmatrix}$ & $\begin{pmatrix}0.680185\\0.0738641+0.612369i\\0.0380878\\0.0738641\end{pmatrix}$\\
\hline
$(0,0,1)^T$ & $\begin{pmatrix}0.765697\\0\\0\\0\end{pmatrix}$ & $\begin{pmatrix}-0.0855116\\0.0738641\\0.0738641\\0.0380878\end{pmatrix}$ & $\begin{pmatrix}0.680185\\0.0738641\\0.0738641\\0.0380878\end{pmatrix}$\\
\end{tabular}
\end{ruledtabular}
\label{tab:currres1}
\end{table}

\section{\label{app:BI}Basic Integrals}
For all calculations, we have used the following definition for the standard scalar integrals:
\begin{align}
H_{mn}(M^2,\bar{m}^2,p^2)=-i\lambda^{4-d}\int\frac{d^dl}{(2\pi)^d}\frac{1}{\left[\bar{m}^2-l^2-i\epsilon\right]^m\left[M^2-(p-l)^2-i\epsilon\right]^n}.
\end{align}
Here, $\bar{m}$ and $M$ are two generic mass parameters, $m$ denotes the number of meson propagators and $n$ represents the number of nucleon propagators. By utilizing Feynman parametrization, the integrals $H_{10}$, $H_{01}$ and $H_{11}$ can be calculated analytically. They take the following form:
\begin{align}
H_{10}(0,\bar{m}^2,0)&=2\bar{m}^2\left[\mathcal{L}+\frac{1}{16\pi^2}\log\left(\frac{\bar{m}}{\lambda}\right)\right],\\
H_{01}(M^2,0,0)&=2M^2\left[\mathcal{L}+\frac{1}{16\pi^2}\log\left(\frac{M}{\lambda}\right)\right],\\
H_{11}^{(0)}(M^2,\bar{m}^2,p^2)&=-2\mathcal{L}+\frac{1}{16\pi^2}+\frac{1}{16\pi^2}\ln{(\lambda^2)}\nonumber\\
       &\quad-\frac{p^2-M^2+\bar{m}^2}{32p^2\pi^2}\ln{(\bar{m}^2)}-\frac{p^2+M^2-\bar{m}^2}{32p^2\pi^2}\ln{(M^2)}\\
       &\quad-\frac{\sqrt{4p^2\bar{m}^2-(p^2-M^2+\bar{m}^2)^2}}{32p^2\pi^2}\arccos{\left(\frac{\bar{m}^2-p^2+M^2}{2\bar{m}M}\right)}.\nonumber
\end{align}
Here we have introduced the quantity
\begin{align}
\mathcal{L}=\frac{1}{16\pi^2}\left(\frac{1}{-\epsilon}+\frac{1}{2}\left(\gamma_E-1-\log{4\pi}\right)\right),
\end{align}
which contains both the $\epsilon$ pole as well as some numerical constants. $\gamma_E$ is the Euler-Mascheroni constant. All tensorial integrals of the form
\begin{align}\label{eq:tensint}
-i\lambda^{4-d}\int\frac{d^dl}{(2\pi)^d}\frac{\left\{l^{\mu},l^{\mu}l^{\nu},\ldots\right\}}{\left[\bar{m}^2-l^2-i\epsilon\right]^m\left[M^2-(p-l)^2-i\epsilon\right]^n},
\end{align}
are calculated using the definitions
\begin{align}
p^{\mu}H^{(1)}_{mn}&=-i\lambda^{4-d}\int\frac{d^dl}{(2\pi)^d}\frac{l^{\mu}}{\left[\bar{m}^2-l^2\right]^m\left[M^2-(p-l)^2\right]^n},\label{eq:LDEC1}\\
g^{\mu\nu}H^{(2)}_{mn}+p^{\mu}p^{\nu}H_{mn}^{(3)}&=-i\lambda^{4-d}\int\frac{d^dl}{(2\pi)^d}\frac{l^{\mu}l^{\nu}}{\left[\bar{m}^2-l^2\right]^m\left[M^2-(p-l)^2\right]^n}.\label{eq:LDEC2}
\end{align}
Thus, the functions $H_{11}^{(1,2,3)}$ and $H_{12}^{(1,2,3)}$ take the following form:
\begin{align}
H^{(1)}_{11}&=\frac{1}{2p^2}\left(H_{10}-H_{01}+(p^2-M^2+\bar{m}^2)H^{(0)}_{11}\right),\\
H^{(2)}_{11}&=\frac{1}{2(d-1)}\left(2\bar{m}^2H_{11}-(p^2-M^2+\bar{m}^2)H_{11}^{(1)}-H_{01}\right),\\
H^{(3)}_{11}&=\frac{1}{2p^2(d-1)}\left((2-d)H_{01}+d(p^2-M^2+\bar{m}^2)H^{(1)}_{11}-2\bar{m}^2H^{(0)}_{11}\right),\\
H^{(1)}_{12}&=\frac{1}{2p^2}\left(H^{(0)}_{11}+(p^2-M^2+\bar{m}^2)H^{(0)}_{12}-H_{02}\right),\\
H^{(2)}_{12}&=\frac{1}{2(1-d)}\left(H_{02}+H_{11}^{(1)}+(p^2-M^2+\bar{m}^2)H_{12}^{(1)}-2\bar{m}^2H^{(0)}_{12}\right),\\
H^{(3)}_{12}&=\frac{1}{2p^2(d-1)}\left((2-d)H_{02}+dH_{11}^{(1)}+d(p^2-M^2+\bar{m}^2)H_{12}^{(1)}-2\bar{m}^2H^{(0)}_{12}\right).
\end{align}
Using the following general properties of the scalar integrals
\begin{align}
-\frac{1}{n}\frac{d}{dM^2}H_{mn}=H_{m,n+1},\qquad -\frac{1}{m}\frac{d}{d\bar{m}^2}H_{mn}=H_{m+1,n},
\end{align}
that are valid for $n\neq0$ and $m\neq0$ respectively, one finds that 
\begin{align}
H^{(\alpha)}_{12}=-\frac{d}{dM^2}H^{(\alpha)}_{11},
\end{align}
where $\alpha=1,2,3$. The same principle can be used to calculate $H_{21}^{(\alpha)}$. When employing infrared regularization~\cite{Becher:1999he} or modified infrared regularization~\cite{Gail:2007} in order to obtain properly renormalized, scale independent results it is neccessary to calculate infrared integrals. After combining the meson propagator with the nucleon propagators using Feynman parametrization, one extends the integration
\begin{align}
\int_0^1dx\rightarrow\int_0^{\infty}dx,
\end{align}
and thus defines the infrared integrals $I_{mn}$. As described in Ref.~\cite{Becher:1999he,Gail:2007}, we can now decompose
\begin{align}
\int_0^{\infty}dx=\int_0^1dx + \int_1^{\infty}dx,
\end{align}
which corresponds to the decomposition $I_{mn}=H_{mn}+R_{mn}$, where $R_{mn}$ are called regulator functions. For the functions containing only meson or nucleon propagators, we define
\begin{align}
I_{m0}=H_{m0},\qquad R_{m0}=0,\qquad I_{0n}=0,\qquad R_{0n}=-H_{0n}.
\end{align}
Additionally, the infrared integral $I_{11}$ and thus the regulator function can be calculated analytically, yielding
\begin{align}
\begin{split}
I_{11}&=\frac{p^2-M^2-\bar{m}^2}{p^2}\mathcal{L}+\frac{p^2-M^2+\bar{m}^2}{32\pi^2p^2}+\frac{p^2-M^2+\bar{m}^2}{16\pi^2p^2}\log{\left(\frac{\lambda}{\bar{m}}\right)}\\
&\quad-\frac{\sqrt{4M^2\bar{m}^2-(\bar{m}^2+M^2-p^2)^2}}{16\pi^2p^2}\arccos{\left(\frac{M^2-p^2-\bar{m}^2}{\sqrt{4\bar{m}^2p^2}}\right)}.
\end{split}
\end{align}
For $p^2=M^2$ one finds that this expression reduces to
\begin{align}
I_{11}=&-\frac{\bar{m}^2}{M^2}\mathcal{L}+\frac{\bar{m}^2}{32\pi^2M^2}-\frac{\bar{m}^2}{16\pi^2M^2}\log{\left(\frac{\bar{m}}{\lambda}\right)}-\frac{\bar{m}\sqrt{1-\frac{\bar{m}^2}{4M^2}}}{8\pi^2M}\arccos{\left(-\frac{\bar{m}}{2M}\right)}.
\end{align}
\section{\label{app:FVF} Finite Volume Functions}
The finite volume functions $H_{10}(\bar{m},L)$, $H_{01}(M,L)$ and $H_{11}^{(0)}(\bar{m},M,p,L)$ for a box of volume $L^3$ have been calculated in Refs.~\cite{Gasser198783,AliKhan:2003cu} and take the following form:
\begin{align}
H_{10}(\bar{m},L)&=\sum_{n=1}^{\infty}\frac{\bar{m}\mu(n)}{4\pi^2\sqrt{n}L}K_1\left(\sqrt{n\bar{m}^2L^2}\right),\label{eq:fvol1}\\
H_{01}(M,L)&=\sum_{n=1}^{\infty}\frac{M\mu(n)}{4\pi^2\sqrt{n}L}K_1\left(\sqrt{nM^2L^2}\right),\label{eq:fvol2}\\
H_{11}^{(0)}(\bar{m},M,p,L)&=\sum_{n=1}^{\infty}\frac{\mu(n)}{8\pi^2}\int_0^{1}dxK_0\left(\sqrt{nL^2\left(x(M^2-(1-x)p^2)+\bar{m}^2(1-x)\right)}\right)\label{eq:fvol3}.
\end{align}
The function $\mu(n)$ represents the multiplicity for each $n=n_1^2+n_2^2+n_3^2$, i.e. the number of different combinations for $n_1$, $n_2$ and $n_3$ that yield the wanted $n$. In our calculation of the finite volume corrections to the magnetic moments we have assumed Lorentz invariance when dealing with tensorial integrals.\\For the calculation of the corrections to the electromagnetic current, we have assumed Lorentz invariance to be broken such that Eqs. \eqref{eq:LDEC1} and \eqref{eq:LDEC2} can no longer be used. The starting point for the calculation of such finite volume functions is again the scalar function 
\begin{align}
I_{\beta}^{\mathbf{A}}\left(\mathcal{M}^2,L\right)&=\left(\frac{1}{L^3}\sum_{\mathbf{k}}-\int\frac{d^3\mathbf{k}}{(2\pi)^3}\right)\frac{1}{\left[\left(\mathbf{k}-\mathbf{A}\right)^2+\mathcal{M}^2\right]^{\beta}},
\end{align}
where $\mathbf{A}$ is an arbitrary three momentum and $\mathcal{M}$ is a generic mass parameter. This function can be calculated by following the descriptions given in Ref.~\cite{Sachrajda:2004mi}. This leads to the following expression:
\begin{align}
I_{\beta}^\mathbf{A}\left(\mathcal{M}^2,L\right)=&\frac{1}{\Gamma(\beta)(4\pi)^{3/2}}\int_0^{\infty}d\alpha\,\alpha^{\beta-5/2}e^{-\alpha\mathcal{M}^2}\left(\prod_{i=1}^{3}\sum_{n_i=-\infty}^{\infty}e^{-\frac{L^2n_i^2}{4\alpha}}\cos{\left(n_i\mathbf{A}_iL\right)}-1\right).
\end{align}
From this expression, the results for the tensorial integrals
\begin{align}
I_{\beta}^{\mathbf{A},i}&=\left(\frac{1}{L^3}\sum_{\mathbf{k}}-\int\frac{d^3\mathbf{k}}{(2\pi)^3}\right)\frac{\mathbf{k}^i}{\left[\left(\mathbf{k}-\mathbf{A}\right)^2+\mathcal{M}^2\right]^{\beta}},\label{eq:tens1}\\
I_{\beta}^{\mathbf{A},ij}&=\left(\frac{1}{L^3}\sum_{\mathbf{k}}-\int\frac{d^3\mathbf{k}}{(2\pi)^3}\right)\frac{\mathbf{k}^i\mathbf{k}^j}{\left[\left(\mathbf{k}-\mathbf{A}\right)^2+\mathcal{M}^2\right]^{\beta}},\label{eq:tens2}
\end{align}
can be derived by simply calculating derivatives with respect to $\mathbf{A}_i$ and $\mathbf{A}_j$. Thus, the results take the form
\begin{align}
I_{\beta}^{\mathbf{A},i}&=\frac{1}{2(\beta-1)}\frac{\partial}{\partial \mathbf{A}_i}I^{\mathbf{A}}_{\beta-1}+\mathbf{A}^iI^{\mathbf{A}}_{\beta},\\
I_{\beta}^{\mathbf{A},ij}&=\frac{1}{4(\beta-1)(\beta-2)}\frac{\partial^2}{\partial \mathbf{A}_i\partial \mathbf{A}_j}I^{\mathbf{A}}_{\beta-2}+ \mathbf{A}^i\mathbf{A}^jI^{\mathbf{A}}_{\beta}\\&\quad+\frac{1}{2(\beta-1)}\left[\mathbf{A}^i\frac{\partial}{\partial \mathbf{A}_j}+\mathbf{A}^j\frac{\partial}{\partial \mathbf{A}_i}+\delta^{ij}\right]I^{\mathbf{A}}_{\beta-1}.\nonumber
\end{align}
This result agrees with the result derived in Ref.~\cite{Tiburzi:2007ep}.
\section{\label{app:CEC}Corrections to the electromagnetic current}
In this section, the results for the recalculation of the Feynman diagrams that contribute to the nucleon matrix element of the isovector electromagnetic current to leading one-loop order in BChPT are collected. The result can be written as,
\begin{align}
\varepsilon_{\mu}\langle N(p',s')|\delta\mathcal{J}^{\mu}_v(L)|N(p,s)\rangle=\bar{u}(p',s')\left[\sum_i\text{Amp}_i(\mathcal{M},L,q,p)\right]u(p,s)\times\eta^{\dagger}\frac{\tau^3}{2}\eta,\label{eq:corrcurr}
\end{align}
where we sum over all Feynman diagrams contributing to $O(p^3)$. Thus, the result is of the following form:
\begin{align}
\text{Amp}_{a}&=\frac{g_A^2}{2f_{\pi}^2}\int_0^1dy\left[H_{10}(m,L)-yM^2I_{3/2}(\mathcal{M}_1^2,y\mathbf{p}')+M\gamma_i\frac{\partial}{\partial (y\mathbf{p}'_i)}I_{1/2}(\mathcal{M}_1^2,y\mathbf{p}')\right]\slashed{\varepsilon},\\
\text{Amp}_{b}&=\frac{g_A^2}{2f_{\pi}^2}\int_0^1dy\,\slashed{\varepsilon}\left[H_{10}(m,L)-yM^2I_{3/2}(\mathcal{M}_1^2,y\mathbf{p})+M\gamma_i\frac{\partial}{\partial (y\mathbf{p}_i)}I_{1/2}(\mathcal{M}_1^2,y\mathbf{p})\right],\\
\begin{split}
\text{Amp}_{c}&=-\frac{1}{4}\left(\text{Amp}_{a}+\text{Amp}_{b}\right)\\
              &\quad+\slashed{\varepsilon}\frac{M^2g_A^2}{8f_{\pi}^2}\int_0^1dx\int_0^1dy\,y\biggl[4I_{3/2}(\mathcal{M}_3^2,\mathbf{p}(x,y))+3\mathcal{M}_3^2I_{5/2}(\mathcal{M}_3^2,\mathbf{p}(x,y))\\
              &\quad+2\mathbf{p}_i(x,y)\frac{\partial}{\partial\mathbf{p}_i(x,y)}I_{3/2}(\mathcal{M}_3^2,\mathbf{p}(x,y))-\frac{\partial^2}{\partial \mathbf{p}_i(x,y)^2}I_{1/2}(\mathcal{M}_3^2,\mathbf{p}(x,y))\biggr]\\
              &\quad-\frac{M^2g_A^2}{4f_{\pi}^2}\int_0^1dx\int_0^1dy\,y\biggl[3\slashed{\varepsilon}\slashed{p}(x,y)I_{1/2}(\mathcal{M}_3^2,\mathbf{p}(x,y))\\
              &\quad-yM\varepsilon_i\frac{\partial}{\partial\mathbf{p}_i(x,y)}I_{3/2}(\mathcal{M}_3^2,\mathbf{p}(x,y))-y(\varepsilon\cdot p) \gamma_i\frac{\partial}{\partial\mathbf{p}_i(x,y)}I_{3/2}(\mathcal{M}_3^2,\mathbf{p}(x,y))\\
              &\quad+\varepsilon_i\gamma_j\frac{\partial^2}{\partial\mathbf{p}_i(x,y)\partial\mathbf{p}_j(x,y)}I_{1/2}(\mathcal{M}_3^2,\mathbf{p}(x,y))\biggr],
\end{split}\\
\text{Amp}_{d}&=-\frac{1}{2f_{\pi}^2}\int_0^1dx\left[\slashed{\varepsilon}I_{1/2}(\mathcal{M}_2^2,x\mathbf{q})+\varepsilon_i\gamma_j\frac{\partial^2}{\partial(x\mathbf{q}_i)\partial(x\mathbf{q}_j)}I_{-1/2}(\mathcal{M}_2^2,x\mathbf{q})\right],\\
\text{Amp}_{e}&=g_A^2\text{Amp}_{d}\nonumber\\
              &\quad-\frac{g_A^2M}{f_{\pi}^2}\int_0^1dy\left[y(\varepsilon\cdot p) I_{3/2}(\mathcal{M}_1^2,y\mathbf{p})-\varepsilon_i\frac{\partial}{\partial (y\mathbf{p}_i)}I_{1/2}(\mathcal{M}_1^2,y\mathbf{p})\right]\nonumber\\
              &\quad+\frac{3g_A^2}{2f_{\pi}^2}M(m^2-q^2)\int_0^1dx\int_0^1dy\,y\biggl[\bar{y}(\varepsilon\cdot p) I_{5/2}(\mathcal{M}_4^2,\mathbf{q}(x,y))\nonumber\\
              &\quad-\varepsilon_i\frac{\partial}{\partial \mathbf{q}_i(x,y)}I_{3/2}(\mathcal{M}_4^2,\mathbf{q}(x,y))\biggr]\\
              &\quad-\frac{g_A^2M}{2f_{\pi}^2}\int_0^1dx\int_0^1dy\,y\biggl[-\slashed\varepsilon I_{3/2}(\mathcal{M}_4^2,\mathbf{q}(x,y))+3\bar{y}(\varepsilon\cdot p) \slashed{q}(x,y)I_{5/2}(\mathcal{M}_4^2,\mathbf{q}(x,y))\nonumber\\
              &\quad-\bar{y}(\varepsilon\cdot p)\gamma_i\frac{\partial}{\partial \mathbf{q}_i(x,y)}I_{3/2}(\mathcal{M}_4^2,\mathbf{q}(x,y))-\slashed{q}(x,y)\varepsilon_i\frac{\partial}{\partial\mathbf{q}_i(x,y)}I_{3/2}(\mathcal{M}_4^2,\mathbf{q}(x,y))\nonumber\\
              &\quad+\varepsilon_i\gamma_j\frac{\partial^2}{\partial \mathbf{q}_i(x,y)\partial \mathbf{q}_j(x,y)}I_{1/2}(\mathcal{M}_4^2,\mathbf{q}(x,y))\biggr]\slashed{q}\nonumber,\\
\text{Amp}_f&=-\frac{1}{f_{\pi}^2}H_{10}(m,L).
\end{align}
In the results above, the following masses
\begin{align}
\mathcal{M}_1^2&=x^2M^2+\bar{x}m^2,\quad\mathcal{M}_2^2=m^2-x\bar{x}q^2,\quad\mathcal{M}_3^2=y^2\left(M^2-x\bar{x}q^2\right)+\bar{y}m^2,\\
\mathcal{M}_4^2&=\bar{y}^2M^2+ym^2-x(1+x)y^2q^2,
\end{align}
and momenta
\begin{align}
p(x,y)&=y(p-qx),\quad q(x,y)=\bar{y}p-xyq,
\end{align}
have been used. They stem from combining the different propagators via Feynman parametrization. The Feynman parameter $x$ has solely been used to combine meson propagators or nucleon propagators, where the parameter $y$ has been used to combine different propagators, e.g. a meson propagator with a nucleon propagator.
\bibliographystyle{apsrev}
\bibliography{bibliography}
\end{document}